\documentclass[aps,prd,preprint,showpacs,groupaddress]{revtex4}

\usepackage{graphicx}

\begin{document}

\title{Tri-bimaximal Neutrino Mixing and Flavor-dependent \\Resonant Leptogenesis}

\author{\bf Zhi-zhong Xing}
\email{xingzz@mail.ihep.ac.cn}

\author{\bf Shun Zhou}
\email{zhoush@mail.ihep.ac.cn}

\affiliation{Institute of High Energy Physics, Chinese Academy of
  Sciences, P.O. Box 918(4), Beijing 100049, China}

\begin{abstract}
We propose a particularly economical neutrino mass model, in which
there are only two right-handed Majorana neutrinos of ${\cal O}(1)$
TeV and their masses are highly degenerate. Its novel
Yukawa-coupling texture together with the seesaw mechanism allows us
to achieve the normal neutrino mass hierarchy with $m^{}_1 = 0$ and
a nearly tri-bimaximal neutrino mixing pattern with the maximal
CP-violating phase: $\theta^{}_{23} = \pi/4$, $|\delta| = \pi/2$ and
$\sin^2 \theta^{}_{12} = (1 - 2 \tan^2 \theta^{}_{13})/3$. One may
also obtain the inverted neutrino mass hierarchy with $m^{}_3 = 0$
and the corresponding neutrino mixing pattern with $\theta^{}_{23}
=\pi/4$ and $\theta^{}_{13} = \delta = 0$. In both cases, it is
possible to interpret the cosmological baryon number asymmetry
$\eta^{}_{\rm B} \approx 6.1 \times 10^{-10}$ through the resonant
leptogenesis mechanism. We demonstrate the significance of
flavor-dependent effects in this leptogenesis scenario: they can
either flip the sign of the flavor-independent prediction for
$\eta^{}_{\rm B}$ in the $m^{}_1 = 0$ case or magnify the magnitude
of the flavor-independent prediction for $\eta^{}_{\rm B}$ about
$50$ times in the $m^{}_3 = 0$ case.
\end{abstract}


\maketitle

\section{Introduction}

The recent solar \cite{SNO}, atmospheric \cite{SK}, reactor
\cite{KM} and accelerator \cite{K2K} neutrino oscillation
experiments have provided us with very robust evidence that
neutrinos are massive and their flavor mixing involves two large
angles ($\theta^{}_{12} \sim 34^\circ$ and $\theta^{}_{23} \sim
45^\circ$) and one small angle ($\theta^{}_{13} < 10^\circ$). These
important results indicate that the minimal standard electroweak
model, in which neutrinos are massless Weyl particles, is
incomplete. A very economical extension of the standard model, the
so-called minimal seesaw model \cite{FGY}, is to introduce two heavy
right-handed Majorana neutrinos ($N^{}_1$ and $N^{}_2$) but preserve
its $SU(2)^{}_{\rm L} \times U(1)^{}_{\rm Y}$ gauge symmetry. In
this case, the Lagrangian relevant for lepton masses can be written
as
\begin{equation}
-{\cal L}^{}_{\rm lepton} = \overline{l^{}_{\rm L}} Y^{}_l E H +
\overline{l^{}_{\rm L}} Y^{}_\nu N H^{\rm c} + \frac{1}{2}
\overline{N^{\rm c}} M^{}_{\rm R} N + {\rm h.c.} \; ,
\end{equation}
where $l^{}_{\rm L}$ denotes the left-handed lepton doublet, $H$ is
the Higgs-boson weak isodoublet ($H^{\rm c} \equiv i\sigma^{}_2
H^*$), $E$ and $N$ stand respectively for the right-handed
charged-lepton and Majorana neutrino singlets. After spontaneous
gauge symmetry breaking, we obtain the charged-lepton mass matrix
$M^{}_l = Y^{}_l v$ and the Dirac neutrino mass matrix $M^{}_{\rm D}
= Y^{}_\nu v$ with $v \approx 174 ~ {\rm GeV}$ being the vacuum
expectation value of $H$. Because $N^{}_1$ and $N^{}_2$ are
$SU(2)^{}_{\rm L}$ singlets, their mass scale is not subject to the
electroweak symmetry breaking and can be much higher than $v$. Then
the seesaw mechanism \cite{SS} works for the effective (light and
left-handed) neutrino mass matrix $M^{}_\nu$; i.e., $M^{}_\nu =
M^{}_{\rm D} M^{-1}_{\rm R} M^T_{\rm D}$ holds as a very good
approximation. The smallness of left-handed neutrino masses $m^{}_i$
(for $i=1,2,3$) is therefore ascribed to the largeness of
right-handed neutrino masses $M^{}_i$ (for $i=1,2$), while the
bi-large neutrino mixing pattern is attributed to a significant
mismatch between the diagonalizations of $M^{}_l$ and $M^{}_\nu$.

Note that ${\cal L}^{}_{\rm lepton}$ allows the
lepton-number-violating decays of $N^{}_i$ (for $i=1,2$) to happen:
$N^{}_i \rightarrow l + H^{\rm c}$ and $N^{}_i \rightarrow l^{\rm c}
+ H$. Since each decay mode can occur at both tree and one-loop
levels, the interference of two decay amplitudes leads to a
CP-violating asymmetry $\varepsilon^{}_i$ between $N^{}_i
\rightarrow l + H^{\rm c}$ and its CP-conjugated process $N^{}_i
\rightarrow l^{\rm c} + H$. If the expansion of the Universe gives
rise to the out-of-equilibrium decays of $N^{}_1$ and (or) $N^{}_2$
\cite{S}, then $\varepsilon^{}_1$ and $\varepsilon^{}_2$ (or one of
them) may result in a net lepton number asymmetry. The latter can
partially be converted into a net baryon number asymmetry via the
nonperturbative sphaleron interaction \cite{Kuzmin}. Such an elegant
baryogenesis-via-leptogenesis mechanism \cite{FY} provides a viable
interpretation of the cosmological baryon number asymmetry,
$\eta^{}_{\rm B} \equiv n^{}_{\rm B}/n^{}_\gamma = (6.1 \pm 0.2)
\times 10^{-10}$, which has recently been extracted from the
three-year WMAP observational data \cite{WMAP}.

The most straightforward test of the seesaw and leptogenesis
mechanisms is to discover the heavy right-handed Majorana neutrinos
and measure the strength of their interactions with other known
particles, for instance, at the Large Hadron Collider (LHC) and the
International Linear Collider (ILC) \cite{Test} in the near or
foreseeable future. Nevertheless, $N^{}_i$ cannot be produced at LHC
unless its mass scale $M^{}_i$ is of ${\cal O}(1)$ TeV up to ${\cal
O}(10)$ TeV, far below the typical seesaw scale $\Lambda^{}_{\rm SS}
\sim 10^{10\cdots 14}$ GeV \cite{SS}. If $M^{}_i \sim {\cal O}(1)$
TeV is taken, for example, a successful leptogenesis to account for
the observed value of $\eta^{}_{\rm B}$ requires that the splitting
between $M^{}_1$ and $M^{}_2$ be extremely small \cite{PU}. This
interesting scenario, usually referred to as the resonant
leptogenesis mechanism, may also get clear of the gravitino
overproduction problem in those supersymmetric extensions of the
standard model with heavy Majorana neutrinos of $M^{}_i \sim
\Lambda^{}_{\rm SS}$. But $m^{}_i \lesssim {\cal O}(1)$ eV and
$M^{}_i \sim {\cal O}(1)$ TeV in the seesaw relation imply that the
corresponding Yukawa couplings must be of ${\cal O}(10^{-6})$ or
smaller, rendering the production and detection of $N^{}_i$ rather
dim at LHC and ILC \cite{PU2}. For this reason, here we shall follow
a less ambitious strategy by leaving out the observability of heavy
right-handed Majorana neutrinos and concentrating on the
phenomenology of neutrino mixing and resonant leptogenesis.

The goal of this paper is to simultaneously explain the observed
neutrino mixing pattern and the cosmological baryon number
asymmetry. We are going to propose a particularly economical and
suggestive seesaw model with two highly degenerate right-handed
Majorana neutrinos of ${\cal O}(1)$ TeV. The novel Yukawa-coupling
texture of our model leads to the normal neutrino mass hierarchy
with $m^{}_1 =0$ and a nearly tri-bimaximal neutrino mixing pattern
with the maximal CP-violating phase: $\theta^{}_{23} = \pi/4$,
$|\delta| = \pi/2$ and $\sin^2 \theta^{}_{12} = (1 - 2 \tan^2
\theta^{}_{13})/3$. On the other hand, one may get the inverted
neutrino mass hierarchy with $m^{}_3 =0$ and the corresponding
neutrino mixing pattern with $\theta^{}_{23} = \pi/4$ and
$\theta^{}_{13} = \delta =0$. We find that it is possible to
establish a straightforward link between CP violation in the decays
of heavy right-handed Majorana neutrinos and that in the
oscillations of light left-handed Majorana neutrinos in the $m^{}_1
=0$ case. As for the cosmological baryon number asymmetry, we show
that the observed value of $\eta^{}_{\rm B}$ can naturally be
interpreted through the resonant leptogenesis mechanism in both
$m^{}_1 =0$ and $m^{}_3 =0$ cases. We point out that the formula of
$\varepsilon^{}_i$ given in Ref. \cite{PU} and that presented in
Ref. \cite{plum} lead to the same numerical result in the model
under discussion, if the parameter space coincides with the
perturbation condition. More interestingly, we demonstrate the
significance of flavor-dependent effects in our resonant
leptogenesis scenario: they can either flip the sign of the
flavor-independent prediction for $\eta^{}_{\rm B}$ in the $m^{}_1 =
0$ case or magnify the magnitude of the flavor-independent
prediction for $\eta^{}_{\rm B}$ about $50$ times in the $m^{}_3 =
0$ case.

\section{Neutrino Mixing}

A salient feature of the minimal seesaw model is that it
automatically predicts $m^{}_1 = 0$ or $m^{}_3 = 0$ \cite{FGY}.
Hence the $3\times 2$ Dirac neutrino mass matrix $M^{}_{\rm D}$ can
be parameterized as
$$
M^{(1)}_{\rm D} \; = \; V^{}_0 \left(\matrix{ 0 & 0 \cr x & 0 \cr 0
& y \cr } \right) U \;
\eqno{\rm (2a)}
$$
with $m^{}_1 = 0$ \cite{KT} or as
$$
M^{(3)}_{\rm D} \; = \; V^{}_0 \left(\matrix{ x & 0 \cr 0 & y \cr 0
& 0 \cr } \right) U \;
\eqno{\rm (2b)}
$$
with $m^{}_3 = 0$, where $V^{}_0$ and $U$ are $3 \times 3$ and $2
\times 2$ unitary matrices, respectively. Without loss of
generality, we adopt the flavor basis in which both the
charged-lepton mass matrix $M^{}_l$ and the right-handed Majorana
neutrino mass matrix $M^{}_{\rm R}$ are diagonal, real and positive.
Then the seesaw relation $M^{}_\nu = M^{}_{\rm D} M^{-1}_{\rm R}
M^T_{\rm D}$ implies that the flavor mixing of light Majorana
neutrinos depends primarily on $V^{}_0$ and the decays of heavy
Majorana neutrinos rely mainly on $U$ \cite{BP}. This
phenomenological observation motivates us to take $V^{}_0$ to be the
tri-bimaximal mixing pattern \cite{TB} \setcounter{equation}{2}
\begin{equation}
V^{}_0 \; = \; \left( \matrix {2/\sqrt{6} & 1/\sqrt{3} & 0 \cr
-1/\sqrt{6} & 1/\sqrt{3} & 1/\sqrt{2} \cr 1/\sqrt{6} & ~~
-1/\sqrt{3} ~~ & 1/\sqrt{2} \cr } \right) \; ,
\end{equation}
which is compatible very well with the best fit of current
experimental data on neutrino oscillations \cite{Vissani}. On the
other hand, the unitary matrix $U$ can be parameterized as
\begin{equation}
U \; = \; \left ( \matrix{ \cos \vartheta & \sin \vartheta \cr -\sin
\vartheta & \cos \vartheta \cr} \right ) \left (
\matrix{e^{-i\alpha} & 0 \cr 0 & e^{+i\alpha} \cr}\right) \; .
\end{equation}
Since $\alpha$ is the only phase parameter in our model, it should
be responsible both for the CP violation in neutrino oscillations
and for the CP violation in $N^{}_i$ decays. In order to implement
the idea of resonant leptogenesis, we assume that two heavy Majorana
neutrino masses are highly degenerate; i.e., the magnitude of $r
\equiv (M^{}_2 - M^{}_1)/M^{}_2$ is strongly suppressed. Indeed $|r|
\sim {\cal O}(10^{-7})$ or smaller has typically been anticipated in
some seesaw models with three right-handed Majorana neutrinos
\cite{PU} to gain the successful resonant leptogenesis.

Given $|r| < {\cal O}(10^{-4})$, the explicit form of $M^{}_\nu$ can
reliably be formulated from the seesaw relation $M^{}_\nu =
M^{}_{\rm D} M^{-1}_{\rm R} M^T_{\rm D}$ by neglecting the tiny mass
splitting between $N^{}_1$ and $N^{}_2$. In such a good
approximation, we obtain
$$
M^{(1)}_\nu \; = \; \frac{y^2}{M^{}_2} \left [ V^{}_0 \left (
\matrix{0 & 0 & 0 \cr 0 & \omega^2 \left(\cos 2\alpha - i\cos
2\vartheta \sin 2\alpha \right)& i \omega \sin 2\vartheta \sin
2\alpha \cr 0 & i \omega \sin 2\vartheta \sin 2\alpha & \cos 2\alpha
+ i\cos 2\vartheta \sin 2\alpha \cr} \right ) V^T_0 \right ] \;
\eqno{\rm (5a)}
$$
for $m^{}_1 = 0$; or
$$
M^{(3)}_\nu \; = \; \frac{y^2}{M^{}_2} \left [ V^{}_0 \left (
\matrix{\omega^2 \left(\cos 2\alpha - i\cos 2\vartheta \sin 2\alpha
\right) & i \omega \sin 2\vartheta \sin 2\alpha & 0 \cr i \omega
\sin 2\vartheta \sin 2\alpha & \cos 2\alpha + i\cos 2\vartheta \sin
2\alpha & 0 \cr 0 & 0 & 0 \cr} \right ) V^T_0 \right ] \;
\eqno{\rm (5b)}
$$
for $m^{}_3 = 0$, where $\omega \equiv x/y$ is defined. One may
follow the procedure outlined in Ref. \cite{KT} to diagonalize the
effective neutrino mass matrix in Eq. (5a) or (5b). For simplicity,
here we fix $\vartheta = \pi/4$ and highlight the role of $\alpha$
in neutrino mixing and leptogenesis. Then both the neutrino mass
spectrum and the neutrino mixing pattern can be achieved from the
simplified version of $M^{(1)}_\nu$ or $M^{(3)}_\nu$.
\begin{itemize}
\item     Normal neutrino mass hierarchy with $m^{}_1 =0$. In this
case, we diagonalize $M^{(1)}_\nu$ by using the transformation
$V^\dagger M^{(1)}_\nu V^* = {\rm Diag} \left \{ 0, m^{}_2, m^{}_3
\right \}$, where $V$ is just the neutrino mixing matrix. After a
straightforward calculation, we have
$$
\begin{array}{rcl}
m^{}_2 & = & \displaystyle \frac{y^2}{2M^{}_2} \left [ \sqrt{\left
(1 + \omega^2 \right )^2 \cos^2 2\alpha + 4 \omega^2 \sin^2 2\alpha}
~ - \left (1 - \omega^2 \right ) \left | \cos 2\alpha \right |
\right ] \; ,
\\
m^{}_3 & = & \displaystyle \frac{y^2}{2M^{}_2} \left [ \sqrt{\left
(1 + \omega^2 \right )^2 \cos^2 2\alpha + 4 \omega^2 \sin^2 2\alpha}
~ + \left (1 - \omega^2 \right ) \left | \cos 2\alpha \right |
\right ] \; ,
\end{array}
\eqno{\rm (6a)}
$$
where $0 < \omega < 1$. Taking account of $m^{}_2 = \sqrt{\Delta
m^2_{21}}$ and $m^{}_3 = \sqrt{\Delta m^2_{21} + |\Delta
m^2_{32}|}$, we obtain $m^{}_2 \approx 8.9 \times 10^{-3}$ eV and
$m^{}_3 \approx 5.1 \times 10^{-2}$ eV by using $\Delta m^2_{21}
\approx 8.0 \times 10^{-5} ~ {\rm eV}^2$ and $|\Delta m^2_{32}|
\approx 2.5 \times 10^{-3} ~ {\rm eV}^2$ \cite{Vissani} as the
typical inputs. Furthermore,
$$
~~ V = \left ( \matrix{2/\sqrt{6} & \cos\theta/\sqrt{3} & i
\sin\theta/\sqrt{3} \cr -1/\sqrt{6} & \cos\theta/\sqrt{3} + i
\sin\theta/\sqrt{2} & \cos\theta/\sqrt{2} + i \sin\theta/\sqrt{3}
\cr 1/\sqrt{6} & -\cos\theta/\sqrt{3} + i \sin\theta/\sqrt{2} &
\cos\theta/\sqrt{2} - i \sin\theta/\sqrt{3} \cr} \right ) \; ,
\;\;\;
\eqno{\rm (7a)}
$$
where $\theta$ is given by $\tan 2\theta = 2 \omega \tan 2\alpha
/(1+\omega^2)$. Comparing this result with the standard
parametrization of $V$ \cite{PDG}, we immediately arrive at
$$
~~\; \sin^2 \theta^{}_{12} = \frac{1 - \sin^2 \theta}{3 - \sin^2
\theta} \; , ~~~\; \sin^2 \theta^{}_{13} = \frac{\sin^2 \theta}{3}
\; , ~~~\; \theta^{}_{23} = \frac{\pi}{4} \; , ~~~\; \delta = -
\frac{\pi}{2} \; , ~~~
\eqno{\rm (8a)}
$$
and vanishing Majorana phases of CP violation. Eq. (8a) implies a
very interesting correlation between $\theta^{}_{12}$ and
$\theta^{}_{13}$: $\sin^2 \theta^{}_{12} = (1 - 2 \tan^2
\theta^{}_{13})/3$. When $\theta^{}_{13} \rightarrow 10^\circ$, we
get $\theta^{}_{12} \rightarrow 34^\circ$, which is almost the
best-fit value of the solar neutrino mixing angle \cite{Vissani}
(i.e., the large-mixing-angle MSW solution to the solar neutrino
problem \cite{MSW}). Note that the smallness of $\theta^{}_{13}$
requires the smallness of $\theta$, or equivalently the smallness of
$\alpha$. Eqs. (6a) and (8a), together with $\theta^{}_{13} <
10^\circ$ and the values of $m^{}_2$ and $m^{}_3$ obtained above,
yield $0.39 \lesssim \omega \lesssim 0.42$, $0^\circ \lesssim \alpha
\lesssim 23^\circ$ and $0^\circ \lesssim \theta \lesssim 18^\circ$.
Eq. (6a) can reliably approximate to $m^{}_2 \approx x^2/M^{}_2$ and
$m^{}_3 \approx y^2/M^{}_2$ for $\alpha \lesssim 10^\circ$. The
Jarlskog rephasing-invariant parameter ${\cal J}$ \cite{J}, which
determines the strength of CP violation in neutrino oscillations, is
found to be $|{\cal J}| = \sin 2\theta/(6\sqrt{6}) \lesssim 0.04$ in
this scenario. It is possible to measure $|{\cal J}| \sim {\cal
O}(10^{-2})$ in the future long-baseline neutrino oscillation
experiments.

\item     Inverted neutrino mass hierarchy with $m^{}_3 =0$. In this
case, we diagonalize $M^{(3)}_\nu$ by using the transformation
$V^\dagger M^{(3)}_\nu V^* = {\rm Diag} \left \{ m^{}_1, m^{}_2, 0
\right \}$. Then we obtain
$$
\begin{array}{rcl}
m^{}_1 & = & \displaystyle \frac{y^2}{2M^{}_2} \left [ \sqrt{\left
(1 + \omega^2 \right )^2 \cos^2 2\alpha + 4 \omega^2 \sin^2 2\alpha}
~ - \left (1 - \omega^2 \right ) \left | \cos 2\alpha \right |
\right ] \; ,
\\
m^{}_2 & = & \displaystyle \frac{y^2}{2M^{}_2} \left [ \sqrt{\left
(1 + \omega^2 \right )^2 \cos^2 2\alpha + 4 \omega^2 \sin^2 2\alpha}
~ + \left (1 - \omega^2 \right ) \left | \cos 2\alpha \right |
\right ] \; ,
\end{array}
\eqno{\rm (6b)}
$$
where $0 < \omega < 1$. Taking account of $m^{}_1 = \sqrt{|\Delta
m^2_{32}| - \Delta m^2_{21}}$ and $m^{}_2 = \sqrt{|\Delta
m^2_{32}|}$, we get $m^{}_1 \approx 4.9 \times 10^{-2}$ eV and
$m^{}_2 \approx 5.0 \times 10^{-2}$ eV by inputting $\Delta m^2_{21}
\approx 8.0 \times 10^{-5} ~ {\rm eV}^2$ and $|\Delta m^2_{32}|
\approx 2.5 \times 10^{-3} ~ {\rm eV}^2$. In addition,
$$
~~\; V = \left ( \matrix{2\cos\theta/\sqrt{6} + i\sin\theta/\sqrt{3}
& \cos\theta/\sqrt{3} + 2i \sin\theta/\sqrt{6} & 0 \cr
-\cos\theta/\sqrt{6} + i\sin\theta/\sqrt{3} & \cos\theta/\sqrt{3} -i
\sin\theta/\sqrt{6} & 1/\sqrt{2} \cr \cos\theta/\sqrt{6} - i
\sin\theta/\sqrt{3} & -\cos\theta/\sqrt{3} + i \sin\theta/\sqrt{6} &
1/\sqrt{2} \cr} \right ) \; , \;\;\;
\eqno{\rm (7b)}
$$
where $\theta$ is given by $\tan 2\theta = 2 \omega \tan
2\alpha/(1+\omega^2)$. When the standard parametrization of $V$
\cite{PDG} is applied to Eq. (7b), we have
$$
~~~ \sin^2 \theta^{}_{12} = \frac{1 + \sin^2 \theta}{3} \; , ~~~~
\theta^{}_{13} = 0 \; , ~~~~ \theta^{}_{23} = \frac{\pi}{4} \; ,
~~~~ \delta = 0 \; , ~~~~ {\cal J} = 0 \; , ~~~
\eqno{\rm (8b)}
$$
and vanishing Majorana phases of CP violation. Eq. (8b) indicates
that $M^{(3)}_\nu$ actually has the $\mu$-$\tau$ symmetry and CP is
conserving at low-energy scales in this scenario. Given $30^\circ <
\theta^{}_{12} < 38^\circ$ \cite{Vissani}, $\theta$ is found to lie
in the range $0 \lesssim \theta \lesssim 22^\circ$. This result,
together with Eq. (6b) and the values of $m^{}_1$ and $m^{}_2$
obtained above, allows us to get $0^\circ \lesssim \alpha \lesssim
22^\circ$ and $0.991 \lesssim \omega \lesssim 0.992$. We observe
that $m^{}_1 \approx x^2/M^{}_2$ and $m^{}_2 \approx y^2/M^{}_2$ are
good approximations of Eq. (6b) for $\alpha \lesssim 10^\circ$.
\end{itemize}
From a phenomenological point of view, the scenario with $m^{}_1 =0$
is more favored and more interesting than the scenario with $m^{}_3
=0$. Both of them can be tested in the near future.

\section{Resonant Leptogenesis}

Now let us consider the lepton-number-violating and CP-violating
decays of two heavy right-handed Majorana neutrinos. Their decay
widths are given by $\Gamma^{}_i = (Y^\dagger_\nu Y^{}_\nu)^{}_{ii}
M^{}_i/(8\pi)$ at the tree level, where $Y^{}_\nu = M^{}_{\rm D}/v$.
With the help of Eq. (2), it is easy to verify that $M^\dagger_{\rm
D} M^{}_{\rm D}$ has a universal form for both $m^{}_1 =0$ and
$m^{}_3 =0$ cases: \setcounter{equation}{8}
\begin{equation}
M^\dagger_{\rm D} M^{}_{\rm D} \; = \; U^\dagger \left ( \matrix{x^2
& 0 \cr 0 & y^2 \cr } \right ) U \; .
\end{equation}
Given the near degeneracy between $M^{}_1$ and $M^{}_2$ as well as
$\vartheta = \pi/4$ for $U$, it turns out that $\Gamma^{}_1 =
\Gamma^{}_2$ is an excellent approximation. But the decay mode
$N^{}_i \rightarrow l + H^{\rm c}$ and its CP-conjugated process
$N^{}_i \rightarrow l^{\rm c} + H$ can actually occur at both tree
and one-loop levels. Hence their CP-violating asymmetry
$\varepsilon^{}_i$, defined as the ratio of $\Gamma (N^{}_i
\rightarrow l + H^{\rm c}) - \Gamma (N^{}_i \rightarrow l^{\rm c} +
H)$ to $\Gamma (N^{}_i \rightarrow l + H^{\rm c}) + \Gamma (N^{}_i
\rightarrow l^{\rm c} + H)$, arises from the interference of two
decay amplitudes \cite{FY}. Note that the self-energy correction
dominates over the vertex correction at the one-loop level, as the
former is resonantly enhanced by the tiny mass splitting between
$N^{}_1$ and $N^{}_2$. In other words, $\varepsilon^{}_i$ primarily
results from the interference between the tree-level amplitude and
its self-energy correction \cite{PU}
\footnote{Here we do not discriminate one lepton flavor from another
in the final states of $N^{}_i$ decays. This flavor-independent
leptogenesis scenario actually serves as a counter example to
illustrate why flavor effects are important and must be taken into
account in a TeV-scale resonant leptogenesis model, as one can
clearly see in section IV.},
\begin{equation}
\varepsilon^{}_{i} \; = \; \frac{{\rm Im} \left [ \left
(Y^\dagger_\nu Y^{}_\nu \right )^2_{ij} \right ]}{\left
(Y^\dagger_\nu Y^{}_\nu \right )^{}_{11} \left (Y^\dagger_\nu
Y^{}_\nu \right )^{}_{22}} \cdot \frac{\left (M^2_i - M^2_j \right )
M^{}_i \Gamma^{}_{j}}{\left (M^2_i - M^2_j \right )^2 + M^2_i
\Gamma^{2}_{j} } \;\; ,
\end{equation}
where $i$ and $j$ run over 1 and 2 but $i\neq j$. Combining Eqs. (9)
and (10), we obtain the explicit expression of $\varepsilon^{}_i$ in
our model:
\begin{equation}
\varepsilon^{}_i = \frac{ - 32 \pi v^2 y^2 \left (1 - \omega^2
\right )^2}{\left (1 + \omega^2 \right ) \left [ 1024 \pi^2 v^4 r^2
+ y^4 \left (1 + \omega^2 \right )^2 \right ]} ~ r \sin 4\alpha \; ,
\end{equation}
in which $r \equiv (M^{}_2 - M^{}_1)/M^{}_2$ has been defined to
describe the mass splitting between two heavy Majorana neutrinos. Of
course, $\varepsilon^{}_1 = \varepsilon^{}_2$ is also an excellent
approximation due to $\vartheta = 45^\circ$. These two CP-violating
asymmetries would vanish if $r =0$ or $\alpha =0$ were taken.

To estimate the order of $\varepsilon^{}_i$ at the TeV scale, we
restrict ourselves to the interesting $\alpha \lesssim 10^\circ$
region and make use of the approximate result $y^2 \approx m^{}_3
M^{}_2$ (or $y^2 \approx m^{}_2 M^{}_2$) obtained above for the
$m^{}_1 =0$ (or $m^{}_3 =0$) case.
\begin{itemize}
\item     In the $m^{}_1 =0$ case, we get $y^2 \approx 5.1 \times
10^{-8} ~ {\rm GeV}^2$ from $m^{}_3 \approx 5.1 \times 10^{-2}$ eV
and $M^{}_2 \approx 1$ TeV. In addition, $\omega \approx 0.42$. Then
Eq. (11) is approximately simplified to
$$
\varepsilon^{}_i \; \approx \; \left \{ \begin{array}{lll} -9.7
\times 10^{-15} ~ r^{-1} \sin 4\alpha \; , & ~~~~ & {\rm for} ~ r
\gg 2.0 \times 10^{-14} \; , \\
-2.5 \times 10^{13} ~ r \sin 4\alpha \; , & & {\rm for} ~ r \ll 2.0
\times 10^{-14} \; , \end{array} \right . ~
\eqno{\rm (12a)}
$$
together with $\varepsilon^{}_i \sim -0.25 \times \sin 4\alpha$ for
$r \sim 2.0 \times 10^{-14}$. Note that $|\varepsilon^{}_i| \sim
{\cal O}(10^{-5})$ is in general expected so as to achieve the
successful leptogenesis (see below). Hence the third possibility $r
\sim {\cal O}(10^{-14})$ requires $\alpha \sim {\cal O}(10^{-4})$,
implying very tiny (unobservable) CP violation in neutrino
oscillations. If $\alpha \sim 5^\circ \cdots 10^\circ$, one may take
either $r\sim 10^{-10}$ or $r \sim 10^{-18}$ to obtain
$|\varepsilon^{}_i| \sim {\cal O}(10^{-5})$.

\item     In the $m^{}_3 =0$ case, we obtain $y^2 \approx 5.0 \times
10^{-8} ~ {\rm GeV}^2$ from $m^{}_2 \approx 5.0 \times 10^{-2}$ eV
and $M^{}_2 \approx 1$ TeV. In addition, $\omega \approx 0.99$. Then
Eq. (11) is approximately simplified to
$$
\varepsilon^{}_i \; \approx \; \left \{ \begin{array}{lll} -3.3
\times 10^{-18} ~ r^{-1} \sin 4\alpha \; , & ~~~~ & {\rm for} ~ r
\gg 3.2 \times 10^{-14} \; , \\
-3.1 \times 10^{9} ~ r \sin 4\alpha \; , & & {\rm for} ~ r \ll 3.2
\times 10^{-14} \; , \end{array} \right . ~
\eqno{\rm (12b)}
$$
together with $\varepsilon^{}_i \sim -5.0 \times 10^{-5} \sin
4\alpha$ for $r \sim 3.2 \times 10^{-14}$. Given either $r \sim
{\cal O}(10^{-14})$ or $r\sim {\cal O}(10^{-13})$, it is
straightforward to arrive at $|\varepsilon^{}_i| \sim {\cal
O}(10^{-5})$ for $\alpha \sim 5^\circ \cdots 10^\circ$. If $r \ll
3.2 \times 10^{-14}$ holds, on the other hand, $|\varepsilon^{}_i|$
will be impossible to reach ${\cal O}(10^{-5})$. Note again that
${\cal J} =0$ has been obtained in Eq. (8b), implying that
$\varepsilon^{}_i \neq 0$ is completely independent of ${\cal J} =0$
in this scenario.
\end{itemize}

As described before, the CP-violating asymmetries $\varepsilon^{}_1$
and $\varepsilon^{}_2$ can give rise to a net lepton number
asymmetry in the Universe, provided its expansion rate is larger
than $\Gamma^{}_1$ and $\Gamma^{}_2$. The nonperturbative sphaleron
interaction may partially convert this lepton number asymmetry into
a net baryon number asymmetry \cite{Buch}, \setcounter{equation}{12}
\begin{equation}
\eta^{}_{\rm B} \approx - 0.96 \times 10^{-2} \sum^{}_i \left
(\kappa^{}_i \varepsilon^{}_i \right ) \approx -1.92 \times 10^{-2}
~ \kappa^{}_i \varepsilon^{}_i \; ,
\end{equation}
where $\kappa^{}_1$ and $\kappa^{}_2$ are the efficiency factors
measuring the washout effects associated with the out-of-equilibrium
decays of $N^{}_1$ and $N^{}_2$, and $\kappa^{}_1 = \kappa^{}_2$ is
an excellent approximation in our model. To evaluate the magnitude
of $\kappa^{}_i$, let us take account of the effective neutrino
masses $\tilde{m}^{}_i \equiv (M^\dagger_{\rm D} M^{}_{\rm
D})^{}_{ii}/M^{}_i = (x^2 + y^2)/(2M^{}_i)$. Of course,
$\tilde{m}^{}_1 = \tilde{m}^{}_2$ is also an excellent
approximation. Given $\alpha \lesssim 10^\circ$,
$$
\tilde{m}^{}_i \approx \frac{1}{2} \left (m^{}_2 + m^{}_3 \right )
\approx 2.9 \times 10^{-2} ~ {\rm eV} \; ~~~~~
\eqno{\rm (14a)}
$$
holds for the normal neutrino mass hierarchy ($m^{}_1 =0$); and
$$
\tilde{m}^{}_i \approx \frac{1}{2} \left (m^{}_1 + m^{}_2 \right )
\approx 5.0 \times 10^{-2} ~ {\rm eV} \; ~~~~~
\eqno{\rm (14b)}
$$
holds for the inverted neutrino mass hierarchy ($m^{}_3 =0$). When
$\tilde{m}^{}_i > 10^{-3} ~ {\rm eV}$, the washout of the baryon
number asymmetry is so effective that the final asymmetry can hardly
depend on the initial one. Alternatively, one may define the
parameters $K^{}_i \equiv \Gamma^{}_i/ H$ at $T = M^{}_i$, where $T$
denotes the temperature of the Universe, $H \equiv 1.66
\sqrt{g^{}_*} T^2/M^{}_{\rm Planck}$ is the Hubble constant with
$g^{}_* \simeq 107$, and $M^{}_{\rm Planck} \approx 1.2\times
10^{19} ~ {\rm GeV}$ represents the Planck mass. The relationship
between $K^{}_i$ and $\tilde{m}^{}_i$ is rather simple: $K^{}_i =
\tilde{m}^{}_i/m^{}_*$, where $m^{}_* \simeq 1.08 \times 10^{-3} ~
{\rm eV}$ is the so-called equilibrium neutrino mass. For
simplicity, the efficiency factors $\kappa^{}_i$ can be estimated by
using the approximate formula \cite{Bari} \setcounter{equation}{14}
\begin{eqnarray}
\kappa^{}_1 \; \approx \; \kappa^{}_2 \; \approx \; \frac{1}{2}
\left(\sum_i K^{}_i \right)^{-1.2} \; ,
\end{eqnarray}
which is valid when the masses of two heavy right-handed Majorana
neutrinos are nearly degenerate and the parameters $K^{}_i$ lie in
the range $5 \lesssim K^{}_i \lesssim 100$. With the help of Eq.
(14), we get $K^{}_i \approx 27$ and $\kappa^{}_i \approx 4.2 \times
10^{-3}$ for the $m^{}_1 =0$ case; or $K^{}_i \approx 46$ and
$\kappa^{}_i \approx 2.2 \times 10^{-3}$ for the $m^{}_3 =0$ case.
Finally, the cosmological baryon number asymmetry is found to be
$$
\eta^{}_{\rm B} \; \approx \; \left \{ \begin{array}{lll} 7.8\times
10^{-19} ~ r^{-1} \sin 4\alpha \; , & ~~~~ & {\rm for} ~ r
\gg 2.0 \times 10^{-14} \;  \\
2.0 \times 10^{9} ~ r \sin 4\alpha \; , & & {\rm for} ~ r \ll 2.0
\times 10^{-14} \;  \end{array} \right .
\eqno{\rm (16a)}
$$
in the $m^{}_1 = 0$ case; or
$$
\eta^{}_{\rm B} \; \approx \; \left \{ \begin{array}{lll} 1.4 \times
10^{-22} ~ r^{-1} \sin 4\alpha \; , & ~~~~ & {\rm for} ~ r
\gg 3.2 \times 10^{-14} \;  \\
1.3 \times 10^{5} ~ r \sin 4\alpha \; , & & {\rm for} ~ r \ll 3.2
\times 10^{-14} \;  \end{array} \right .
\eqno{\rm (16b)}
$$
in the $m^{}_3 = 0$ case.

Note that the numerical results in Eq. (16) are obtained by taking
$M^{}_2 \approx 1$ TeV. Other numerical results can similarly be
achieved by starting from Eq. (11) and allowing $M^{}_2$ to vary,
for instance, from 1 TeV to 10 TeV. It is certainly possible to get
$\eta^{}_{\rm B} \approx 6.1 \times 10^{-10}$ from Eq. (16) with a
proper choice of $r$ and $\alpha$, either in the $m^{}_1 =0$ case or
in the $m^{}_3 =0$ case. But one should keep in mind that the
flavor-independent leptogenesis scenario under discussion has a
potential problem, because it ignores the fact that all the Yukawa
interactions of charged leptons are in thermal equilibrium at
temperatures of ${\cal O}(1)$ TeV \cite{Barbieri}. We shall consider
the flavor-dependent effect in the next section and demonstrate its
significance in our resonant leptogenesis model.


We remark that the formula of $\varepsilon^{}_i$ in Eq. (10) is
quoted from Ref. \cite{PU}. A different formula has recently been
presented in Ref. \cite{plum} to describe the resonant CP-violating
asymmetry for two nearly degenerate right-handed Majorana neutrinos:
\setcounter{equation}{16}
\begin{eqnarray}
\varepsilon^\prime_i \; = \; \frac{{\rm Im} \left [ \left
(Y^\dagger_\nu Y^{}_\nu \right )^2_{ij} \right ]}{\left
(Y^\dagger_\nu Y^{}_\nu \right )^{}_{11} \left (Y^\dagger_\nu
Y^{}_\nu \right )^{}_{22}} \cdot \frac{\left (M^2_i - M^2_j \right )
M^{}_i \Gamma^{}_{j}}{\left (M^2_i - M^2_j \right )^2 + \left(M^{}_j
\Gamma^{}_{j} - M^{}_i \Gamma^{}_{i}\right)^2 } \;\; ,
\end{eqnarray}
where the perturbation condition $(M^{}_2 - M^{}_1)/M^{}_2 \gg {\rm
max}[(Y^\dagger_\nu Y^{}_\nu)^{}_{ij}/(16\pi^2)]$ should be
satisfied. Which formula is more reliable remains controversial at
present. Here let us compare between the results of
$\varepsilon^{}_i$ and $\varepsilon^\prime_i$ in our model. Taking
account of
\begin{eqnarray}
M^{}_j \Gamma^{}_j - M^{}_i \Gamma^{}_i \; = \; \hat{r} \left (M^2_j
- M^2_i \right ) \; ,
\end{eqnarray}
where $\hat{r} \equiv \Gamma^{}_i/M^{}_i = \Gamma^{}_j/M^{}_j =
y^2(1+\omega^2)/(16\pi v^2)$ for $i\neq j$, we obtain the ratio
\begin{eqnarray}
\frac{\varepsilon^\prime_i}{\varepsilon^{}_i} \; =\; \frac{\left
(M^2_i - M^2_j \right )^2 + M^2_i \Gamma^2_{j}}{\left (M^2_i - M^2_j
\right )^2 + \left(M^{}_j \Gamma^{}_{j} - M^{}_i
\Gamma^{}_{i}\right)^2 } \; = \; \frac{1 + \hat{r}^2/\left( 4r^2
\right)}{1+ \hat{r}^2} \;
\end{eqnarray}
from Eqs. (10) and (17) as an excellent approximation. Note that
$\hat{r}/2 \approx 2.0 \times 10^{-14}$ (for $m^{}_1 = 0$) or
$\hat{r}/2 \approx 3.2 \times 10^{-14}$ (for $m^{}_3 = 0$) in the
numerical examples taken above. Note also that the perturbation
condition used in deriving Eq. (17) is equivalent to $r\gg
\hat{r}/(2\pi)$. Hence we arrive at $\varepsilon^\prime_i \approx
\varepsilon^{}_i$ when $r \gg \hat{r}/2$ holds, or equivalently
$r\gg \hat{r}/(2\pi)$ is satisfied. In other words, the formula of
$\varepsilon^{}_i$ given in Ref. \cite{PU} and that presented in
Ref. \cite{plum} lead to the same numerical results in our
leptogenesis scenario, if and only if the parameter space coincides
with the perturbation condition. This numerical agreement is by no
means a hint that there might not exist any serious discrepancy
between Refs. \cite{PU} and \cite{plum}. It is actually desirable to
understand the properties of unstable particles in quantum field
theories to the utmost extent, so as to clarify the potential
ambiguities associated with $\varepsilon^{}_i$ and
$\varepsilon^\prime_i$ (or one of them) in the resonant leptogenesis
models.

\section{Flavor Effects}

Flavor-dependent effects in leptogenesis have recently attracted a
lot of attention \cite{Flavor}. Since all the Yukawa interactions of
charged leptons are in thermal equilibrium at the TeV scale, the
flavor issues of $N^{}_i$ decays should be taken into account in our
model. In the framework of resonant leptogenesis, it is
straightforward to calculate the CP-violating asymmetry between
$N^{}_i \rightarrow l^{}_\alpha + H^{\rm c}$ and $N^{}_i \rightarrow
l^{\rm c}_\alpha + H$ decays for each lepton flavor $\alpha$ ($=e$,
$\mu$ or $\tau$):
\begin{eqnarray}
\varepsilon^{}_{i \alpha} & \equiv & \frac{\Gamma \left(N^{}_i
\rightarrow l^{}_\alpha + H^{\rm c}\right) - \Gamma \left(N^{}_i
\rightarrow l^{\rm c}_\alpha + H\right)}{\displaystyle \sum_\alpha
\left[\Gamma \left(N^{}_i \rightarrow l^{}_\alpha + H^{\rm c}\right)
+ \Gamma \left(N^{}_i \rightarrow
l^{\rm c}_\alpha + H\right)\right]} \nonumber \\
&=& \frac{8\pi \left(M^2_i - M^2_j\right) {\rm Im}\left\{
\left(Y^{}_\nu\right)^{}_{\alpha j} \left(Y^{}_{\nu
}\right)^{*}_{\alpha i} M^{}_i \left[M^{}_i \left(Y^\dagger_\nu
Y^{}_\nu \right)^{}_{ji} + M^{}_j \left(Y^\dagger_\nu Y^{}_\nu
\right)^{}_{ij} \right] \right\}}{\left[ 64\pi^2 \left(M^2_i
-M^2_j\right)^2 + M^4_i \left(Y^\dagger_\nu Y_\nu \right)^{2}_{jj}
\right]\left(Y^\dagger_\nu Y_\nu \right)^{}_{ii}} \; ,
\end{eqnarray}
where $i$ and $j$ run over $1$ and $2$ but $i \neq j$. Combining
Eqs. (2) and (20), we explicitly obtain
$$
\varepsilon^{}_{i e} \; = \; \frac{\omega^2}{3 \left(\omega^2
-1\right)} ~ \varepsilon^{}_i \; , ~~~~~~~~~~ \varepsilon^{}_{i \mu}
\; = \; \varepsilon^{}_{i \tau} \; = \; \frac{2\omega^2 - 3}{6
\left(\omega^2 -1\right)} ~ \varepsilon^{}_i \;
\eqno{\rm (21a)}
$$
in the $m^{}_1 = 0$ case; and
$$
\varepsilon^{}_{i e} \; = \; \frac{2\omega^2 - 1}{3 \left(\omega^2
-1\right)} ~ \varepsilon^{}_i \; , ~~~~~~~~~~ \varepsilon^{}_{i \mu}
\; = \; \varepsilon^{}_{i \tau} \; = \; \frac{\omega^2 - 2}{6
\left(\omega^2 -1\right)} ~ \varepsilon^{}_i \;
\eqno{\rm (21b)}
$$
in the $m^{}_3 = 0$ case, where $\varepsilon^{}_i$ has been given in
Eq. (11). Because $\varepsilon^{}_1 = \varepsilon^{}_2$ is an
excellent approximation, one can see from Eq. (21) that
$\varepsilon^{}_{1 \alpha} = \varepsilon^{}_{2 \alpha}$ is also an
excellent approximation. In view of $0.39 \lesssim \omega \lesssim
0.42$ for $m^{}_1 =0$ or $0.991 \lesssim \omega \lesssim 0.992$ for
$m^{}_3 = 0$, we have $\varepsilon^{}_{ie} <0$ and
$\varepsilon^{}_{i \mu} = \varepsilon^{}_{i \tau} >0$ in both cases.
The sum of these three flavor-dependent asymmetries is just the
total CP-violating asymmetry $\varepsilon^{}_i$; i.e.,
$\varepsilon^{}_{ie} + \varepsilon^{}_{i\mu} +
\varepsilon^{}_{i\tau} = \varepsilon^{}_i$ holds exactly.

Once the initial values of $\varepsilon^{}_{i\alpha}$ are fixed, the
final result of $\eta^{}_{\rm B}$ will be governed by a set of
flavor-dependent Boltzmann equations including the (inverse) decay
and scattering processes as well as the nonperturbative sphaleron
interaction \cite{PU2,Barbieri,Flavor}. In order to estimate the
washout effects, one may introduce the parameters
\setcounter{equation}{21}
\begin{eqnarray}
K^{}_{i \alpha} \; = \; \frac{\Gamma \left(N^{}_i \rightarrow
l^{}_\alpha + H^{\rm c}\right) + \Gamma \left(N^{}_i \rightarrow
l^{\rm c}_\alpha + H\right)}{\displaystyle \sum_\alpha \left[\Gamma
\left(N^{}_i \rightarrow l^{}_\alpha + H^{\rm c}\right) + \Gamma
\left(N^{}_i \rightarrow l^{\rm c}_\alpha + H\right)\right]} K^{}_i
= \frac{\left| (Y^{}_\nu)^{}_{\alpha i} \right|^2}
{\left(Y^\dagger_\nu Y_\nu \right)^{}_{ii}} K^{}_i \; ,
\end{eqnarray}
in which $K^{}_i \equiv \Gamma^{}_i/H$ at $T=M^{}_i$ has been
defined before. Of course, $K^{}_{ie} + K^{}_{i\mu} + K^{}_{i\tau} =
K^{}_i$ holds. With the help of Eq. (2), we get either
$$
\begin{array}{rcl}
K^{}_{1 e} & = & K^{}_{2 e} = \frac{\displaystyle
\omega^2}{\displaystyle 3 \left(1 + \omega^2\right)}K^{}_i \; ,
\\
K^{}_{1 \mu} & = & K^{}_{2 \tau} = \frac{\displaystyle 2\omega^2 -
2\sqrt{6} ~ \omega +3}{\displaystyle 6 \left(1+\omega^2\right)}
K^{}_i \; ,
\\
K^{}_{1 \tau} & = & K^{}_{2 \mu} = \frac{\displaystyle 2\omega^2 +
2\sqrt{6} ~ \omega +3}{\displaystyle 6 \left(1+\omega^2\right)}
K^{}_i \; ,
\end{array}
\eqno{\rm (23a)}
$$
in the $m^{}_1 = 0$ case; or
$$
\begin{array}{rcl}
K^{}_{1 e} & = & \frac{\displaystyle 1 - 2\sqrt{2} \omega +
2\omega^2}{\displaystyle 3 \left(1 + \omega^2\right)}K^{}_i \; , \\
K^{}_{1 \mu} & = & K^{}_{1 \tau} = \frac{\displaystyle 2 + 2\sqrt{2}
\omega + \omega^2}{\displaystyle 6 \left(1 +
\omega^2\right)}K^{}_i \; , \\
K^{}_{2 e} & = & \frac{\displaystyle 1 + 2\sqrt{2} \omega +
2\omega^2}{\displaystyle 3 \left(1 + \omega^2\right)}K^{}_i \; , \\
K^{}_{2 \mu} &=& K^{}_{2 \tau} = \frac{\displaystyle 2 - 2\sqrt{2}
\omega + \omega^2}{\displaystyle 6 \left(1 + \omega^2\right)}K^{}_i
\; ,
\end{array}
\eqno{\rm (23b)}
$$
in the $m^{}_3 = 0$ case.  The flavor-dependent effects in
leptogenesis with almost degenerate heavy Majorana neutrinos have
been elucidated in Ref. \cite{Bari}, where a useful approximation
for the efficiency factor of each flavor is given as
\setcounter{equation}{23}
\begin{eqnarray}
\kappa^{}_{i \alpha} \; \approx \; \frac{2}{K^{}_\alpha z^{}_{\rm
B}(K^{}_\alpha)}\left[1-\exp\left(-\frac{K^{}_\alpha z^{}_{\rm
B}(K^{}_\alpha)}{2}\right)\right] \;
\end{eqnarray}
with $z^{}_{\rm B}(K^{}_\alpha) \simeq 2+4 K^{0.13}_{\alpha}
\exp\left(-2.5/K^{}_\alpha\right)$ and $K^{}_\alpha = K^{}_{1
\alpha} + K^{}_{2\alpha}$. The above expression can well approximate
to the power law $\kappa^{}_{i \alpha} \simeq 0.5/K^{1.2}_\alpha$
for $5 \lesssim K^{}_\alpha \lesssim 100$. Recalling $0.39 \lesssim
\omega \lesssim 0.42$ (or $0.991 \lesssim \omega \lesssim 0.992$)
and $K^{}_i \approx 27$ (or $K^{}_i \approx 46$) for $m^{}_1 =0$ (or
$m^{}_3 =0$), one can then evaluate the decay parameters
$K^{}_\alpha$: $K^{}_e \approx 2.4$ and $K^{}_\mu = K^{}_\tau
\approx 25.8$ (or $K^{}_e \approx 46$ and $K^{}_\mu = K^{}_\tau
\approx 23$). The final baryon number asymmetry can be calculated
via
\begin{eqnarray}
\eta^{\rm f}_{\rm B} \; \approx \; -0.96 \times 10^{-2} \sum_i
\sum_\alpha \left (\varepsilon^{}_{i \alpha} \kappa^{}_{i \alpha}
\right) \; ,
\end{eqnarray}
where the superscript means that flavor effects have been included;
$\varepsilon^{}_{i \alpha}$ and $\kappa^{}_{i \alpha}$ are given in
Eqs. (21) and (24), respectively. A straightforward comparison
between $\eta^{}_{\rm B}$ in Eq. (13) and $\eta^{\rm f}_{\rm B}$ in
Eq. (25) can clearly demonstrate the significance of flavor effects
in our resonant leptogenesis scenario:
\begin{eqnarray}
\frac{\eta^{\rm f}_{\rm B}}{\eta^{}_{\rm B}} \; = \;
\frac{\displaystyle \sum_i \sum_\alpha \left (\varepsilon^{}_{i
\alpha} \kappa^{}_{i \alpha} \right)}{\displaystyle \sum_i
\left(\varepsilon_i \kappa^{}_i\right)} \; \approx \; \left \{
\matrix{-0.71 ~~~~~ (m^{}_1 =0) \; , ~ \cr 52.1 ~~~~~~ (m^{}_3 =0)
\; . \cr} \right . \;
\end{eqnarray}
Note that flavor effects flip the sign of $\eta^{}_{\rm B}$ in the
$m^{}_1=0$ case. This nontrivial result can be understood as
follows. The novel Yukawa structure of our model dictates the
electron flavor to make a negative contribution to the CP-violating
asymmetry ($\varepsilon^{}_{i e} < 0$) and to have a large
efficiency factor (due to the smallness of $K^{}_e$; i.e., $K^{}_e
\approx 2.4$). Although the other two flavors contribute
significantly to the CP-violating asymmetry, their corresponding
washout effects are much stronger ($K^{}_\mu = K^{}_\tau \approx
25.8$). Hence the overall effects of three flavors in $\eta^{\rm
f}_{\rm B}$ make its sign opposite to that of $\eta^{}_{\rm B}$ in
the $m^{}_1 =0$ case. In the $m^{}_3 =0$ case, the large and
negative contribution of the electron flavor to the CP-violating
asymmetry is badly diluted by its extremely small efficiency factor
(due to the largeness of $K^{}_e$; i.e., $K^{}_e \approx 46$), and
thus the muon and tau flavors make the dominant contributions to
$\eta^{\rm f}_{\rm B}$ via both $\varepsilon^{}_{i\mu} =
\varepsilon^{}_{i\tau}$ and $\kappa^{}_{i\mu} = \kappa^{}_{i\tau}$.
That is why $\eta^{\rm f}_{\rm B}$ has the same sign as
$\eta^{}_{\rm B}$ in the $m^{}_3 =0$ case, in which the large ratio
$\eta^{\rm f}_{\rm B}/\eta^{}_{\rm B} \approx 52.1$ is essentially
ascribed to the large enhancement factors
$\varepsilon^{}_{i\mu}/\varepsilon^{}_i \sim 10$ and $(K^{}_1 +
K^{}_2)/K^{}_\mu \sim 4$.

Combining Eqs. (16) and (26), we are then able to obtain the
observed value of $\eta^{\rm f}_{\rm B}$ via the flavor-dependent
TeV-scale leptogenesis with proper inputs of $r$ and $\alpha$. As an
example, FIG. 1 illustrates the numerical correlation between $r$
and $\alpha$ to get $\eta^{\rm f}_{\rm B} = 6.1 \times 10^{-10}$ in
both $m^{}_1 =0$ and $m^{}_3=0$ cases, where $M^{}_2 =1$ TeV, 2 TeV,
3 TeV, 4 TeV and 5 TeV have typically been input and Eq. (11) has
also been used. Note that we only concentrate on the region $r \gg
{\cal O}(10^{-14})$, which satisfies the perturbation condition
given in Ref. \cite{plum}. One may clearly see the similar behavior
of $r$ changing with $\alpha$ in both cases, in which $\eta^{\rm
f}_{\rm B} \propto \varepsilon^{}_i \propto y^2 r^{-1} \sin 4\alpha
\propto M^{}_2 r^{-1} \sin 4\alpha$ holds as the leading-order
approximation. In order to get the correct sign for the cosmological
baryon number asymmetry, however, we should take $-10^\circ < \alpha
< 0^\circ$ in the $m^{}_1=0$ case. Since negative $\alpha$ leads to
negative $\theta$ in Eq. (7a), the sign of $\delta$ in Eq. (8a) will
flip. But this modification does not change the results of neutrino
masses and flavor mixing angles obtained in Eqs. (6a) and (8a).

\section{Further discussions}

In summary, we have proposed a very simple but suggestive seesaw
model with two highly degenerate right-handed Majorana neutrinos of
${\cal O}(1)$ TeV. Its novel Yukawa-coupling texture leads to the
normal neutrino mass hierarchy with $m^{}_1 =0$ and a nearly
tri-bimaximal neutrino mixing pattern with the maximal CP-violating
phase: $\theta^{}_{23} = \pi/4$, $|\delta| = \pi/2$ and $\sin^2
\theta^{}_{12} = (1 - 2 \tan^2 \theta^{}_{13})/3$. On the other
hand, it is possible to get the inverted neutrino mass hierarchy
with $m^{}_3 =0$ and the corresponding neutrino mixing pattern with
$\theta^{}_{23} = \pi/4$ and $\theta^{}_{13} = \delta =0$. A
straightforward link between CP violation in the decays of heavy
right-handed Majorana neutrinos and that in the oscillations of
light left-handed Majorana neutrinos can be established in the
$m^{}_1 =0$ case. We have successfully interpreted the cosmological
matter-antimatter asymmetry $\eta^{}_{\rm B} \approx 6.1 \times
10^{-10}$ through the flavor-dependent resonant leptogenesis
mechanism in both $m^{}_1 =0$ and $m^{}_3 =0$ cases. In particular,
we have shown that flavor effects can either flip the sign of the
flavor-independent prediction for $\eta^{}_{\rm B}$ in the $m^{}_1 =
0$ case or magnify the magnitude of the flavor-independent
prediction for $\eta^{}_{\rm B}$ about 50 times in the $m^{}_3 = 0$
case.

Some more discussions, remarks and comments on our model are in
order.
\begin{itemize}
\item     We have made three assumptions in building the model.
Taking $V^{}_0$ to be the tri-bimaximal mixing pattern is purely a
phenomenological assumption \cite{Kang}, so is taking $U$ to be the
maximal mixing pattern with a CP-violating phase. Both of them might
be able to result from certain flavor symmetries (e.g., discrete
$A^{}_4$ and $Z^{}_2$ symmetries \cite{MS}). The third assumption is
the near degeneracy between $M^{}_1$ and $M^{}_2$, or equivalently
the smallness of $r$. It seems a bit contrived to take such tiny
values of $r$, as shown in FIG. 1, to achieve the successful
baryogenesis via leptogenesis. Why is $r$ so small? A few authors
\cite{Joaquim} have conjectured that $M^{}_1$ and $M^{}_2$ might be
exactly degenerate at an energy scale higher than the seesaw scale
(e.g., the GUT scale) and their small splitting at the seesaw scale
is attributed to the radiative corrections.

\item     The minimal seesaw model with two heavy right-handed Majorana
neutrinos $N^{}_1$ and $N^{}_2$ can be regarded as a limiting
version of the conventional seesaw model with three right-handed
Majorana neutrinos $N^{}_1$, $N^{}_2$ and $N^{}_3$, in which
$N^{}_3$ is so heavy that it decouples from the theory at the very
early stage. In order to gain the special mass spectrum (i.e.,
$M^{}_1 \approx M^{}_2 \ll M^{}_3$), one possibility is to make use
of the Froggatt-Nielsen (FN) mechanism and assign the charges of
three right-handed Majorana neutrinos as $Q^{}_{\rm FN}(N^{}_{\rm 1
R}) =-1$, $Q^{}_{\rm FN}(N^{}_{\rm 2 R})=+1$ and $Q^{}_{\rm
FN}(N^{}_{\rm 3 R})=0$ \cite{PU}. Such a scenario allows $N^{}_1$
and $N^{}_2$ to be of ${\cal O}(1)$ TeV and $N^{}_3$ to be close to
the GUT scale, and it can also accommodate the Yukawa couplings of
${\cal O}(10^{-7})$. Besides flavor symmetries, the supersymmetry
breaking may do the same job, as shown in Ref. \cite{Hambye}.

\item     Radiative corrections to the light neutrino masses,
the mixing angles and the CP-violating phases should also be taken
into account, when those parameters run from the seesaw scale down
to the electroweak scale. A generic analysis of the
renormalization-group running effects in the minimal seesaw model
has been done in Ref. \cite{Mei}. Because $M^{}_1 \approx M^{}_2
\sim {\cal O}(1)$ TeV is specified in our model, however, all of
such running effects are expected to be negligibly small. Even in
the supersymmetric case with large $\tan\beta$, significant
radiative corrections cannot emerge between the energy scales of
${\cal O}(1)$ TeV and the electroweak scale.

\item     We stress that this TeV-scale neutrino mass model is also
viable in the framework of the minimal supersymmetric standard
model. In particular, the above calculation of $\eta^{}_{\rm B}$ can
simply be extended to accommodate supersymmetry. The size of
$\varepsilon^{}_i$ in the supersymmetric case is twice as large as
in the standard model, so is the magnitude of $g^{}_*$ \cite{Buch}.
The enhancement of $g^{}_*$ gives rise to the same order suppression
of the ${\cal O}(10^{-2})$ dilution coefficient in Eq. (13) or Eq.
(25), hence the estimate of $\eta^{}_{\rm B}$ is not significantly
changed by the introduction of supersymmetry. This observation,
which has been confirmed by our detailed analysis, implies that the
numerical results shown in FIG. 1 are expected to be roughly
(order-of-magnitude) valid for the minimal supersymmetric standard
model.
\end{itemize}

The upcoming neutrino oscillation experiments will test the
predictions of our model for the neutrino mass spectrum, flavor
mixing angles and CP violation. A more intriguing test of such
seesaw-plus-leptogenesis models is certainly the direct search for
TeV-scale right-handed Majorana neutrinos
\footnote{The predictions for possible signatures of heavy
right-handed Majorana neutrinos $N^{}_i$ are strongly
model-dependent. For example, they inevitably rely on the mixing
strength between $N^{}_i$ and three charged leptons. The latter has
been constrained to some extent by current experimental data on the
lepton universality and lepton-flavor-violating processes. See Ref.
\cite{Test} for detailed discussions.},
which can be done at LHC and ILC in the future. Due to the tiny
Yukawa couplings restricted by the models themselves, however, the
observability of such new particles will be a big experimental
challenge.

\begin{acknowledgments}
This work was supported in part by the National Natural Science
Foundation of China.
\end{acknowledgments}

\begin{figure*}[]
\begin{center}
\includegraphics[scale=0.85, bb=50 20 700 800 ]{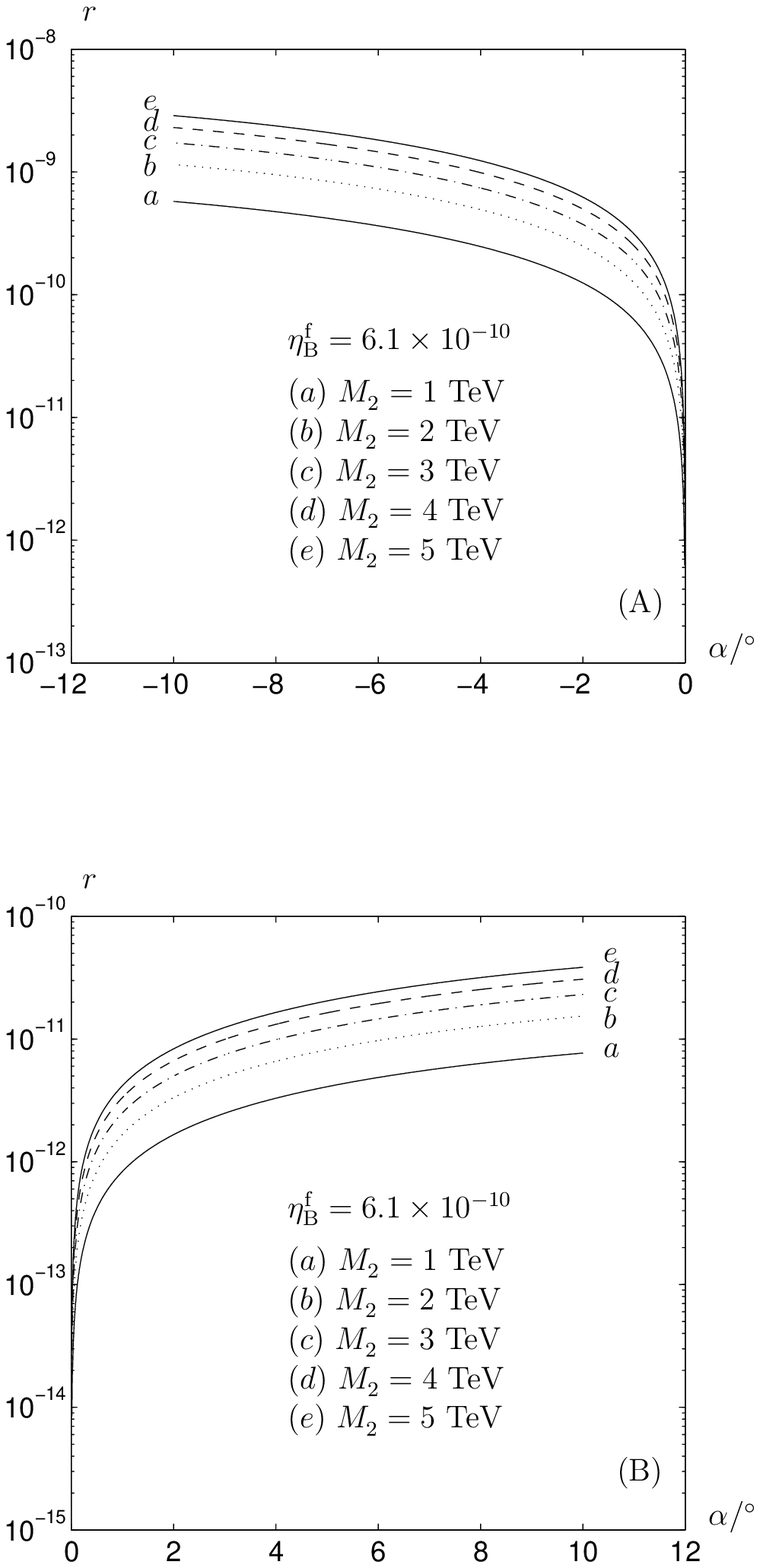}
\vspace{-5cm} \caption{ Numerical illustration of the correlation
between the mass splitting parameter $r$ and the CP-violating phase
$\alpha$ to achieve the successful flavor-dependent resonant
leptogenesis at the TeV scale: (A) in the $m^{}_1 = 0$ case, and (B)
in the $m^{}_3 = 0$ case.}
\end{center}
\end{figure*}

\end{document}